\documentclass[prd,aps,twocolumn, showpacs, nofootinbib,superscriptaddress,notitlepage]{revtex4-1}
\usepackage{amssymb,amsthm,amsmath}
\usepackage{graphicx}   
\usepackage{color}      
\usepackage{slashed}    
\usepackage{verbatim}
\usepackage[normalem]{ulem}
\usepackage{rotating}   
\usepackage{multirow}   

\begin{document}

\title{A comparative lattice analysis of $SU(2)$ dark glueball}

\author{Min-Huan Chu}
\affiliation{INPAC, Key Laboratory for Particle Astrophysics and Cosmology (MOE),  Shanghai Key Laboratory for Particle Physics and Cosmology, School of Physics and Astronomy, Shanghai Jiao Tong University, Shanghai 200240, China}
\affiliation{Yang Yuanqing Scientific Computering Center, 
Tsung-Dao Lee Institute, Shanghai Jiao Tong University, Shanghai 200240, China}

\author{Jun-Hui Lai}
\email{Corresponding author: lai.junh@sjtu.edu.cn}
\affiliation{INPAC, Key Laboratory for Particle Astrophysics and Cosmology (MOE),  Shanghai Key Laboratory for Particle Physics and Cosmology, School of Physics and Astronomy, Shanghai Jiao Tong University, Shanghai 200240, China}

\author{Wei Wang}
\email{Corresponding author: wei.wang@sjtu.edu.cn}
\affiliation{INPAC, Key Laboratory for Particle Astrophysics and Cosmology (MOE),  Shanghai Key Laboratory for Particle Physics and Cosmology, School of Physics and Astronomy, Shanghai Jiao Tong University, Shanghai 200240, China}
\affiliation{Southern Center for Nuclear-Science Theory (SCNT), Institute of Modern Physics, Chinese Academy of Sciences, Huizhou 516000, Guangdong Province, China}

\author{Jialu Zhang}
\affiliation{INPAC, Key Laboratory for Particle Astrophysics and Cosmology (MOE),  Shanghai Key Laboratory for Particle Physics and Cosmology, School of Physics and Astronomy, Shanghai Jiao Tong University, Shanghai 200240, China}

\author{Qianteng Zhu}
\affiliation{INPAC, Key Laboratory for Particle Astrophysics and Cosmology (MOE),  Shanghai Key Laboratory for Particle Physics and Cosmology, School of Physics and Astronomy, Shanghai Jiao Tong University, Shanghai 200240, China}

\begin{abstract}
We study the mass and scattering cross section of $SU(2)$ glueballs as dark matter candidates using lattice simulations. We employ both naive and improved $SU(2)$ gauge actions in $3+1$ dimensions with several $\beta$ values, and   adopt  both the traditional  Monte Carlo method and the flow-based model based on  machine learning techniques to generate lattice configurations. The mass of dark scalar glueball  with $J^{PC}=0^{++}$ and the NBS wave function are calculated.  
Using a coupling constant of $\beta=2.2$ as an illustration, we compare the dark glueball mass calculated from the configurations generated from the two methods. While consistent results can be achieved, the two methods demonstrate distinct advantages.  Using the Runge-Kutta method, we extract the glueball interaction potential and  two-body scattering cross section.  From the observational constraints, we obtain the lower bound of the mass of scalar glueball as candidates of dark matter. 
\end{abstract}

\maketitle
\section{Introduction}
Dark matter (DM) that gives rise to about a quarter of the universe's mass remains one of the most mysterious objects in particle physics. Primarily recognized through its gravitational influence on galaxies and cosmic structures, it   might be also undetectable through electromagnetic means.
The weakly interacting massive particles miracle  (WIMP), the suggestive coincidence of the weak coupling DM with a proper  density that would have thermally been generated, motivated  experimental tests through their direct scattering,  decays to visible cosmic rays, and DM productions at collider experiments, but no positive result has been reported so far~(for an incomplete list on this topics, please see Refs.~\cite{Bai:2010hh,Fox:2011fx,PandaX-II:2016vec,Slatyer:2016qyl,LUX:2016ggv,ATLAS:2017bfj,PandaX-II:2017hlx,PICO:2017tgi,XENON:2018voc,ATLAS:2021kxv,PandaX-4T:2021bab,Foster:2021ngm,PandaX:2023toi,PandaX:2022ood,XENON:2023cxc}).

Aside from  WIMP, several other interesting DM models exist. An interesting scenario suggests dark matter is likely to be  glueballs,  particles composed of strongly interacting  dark gluons bound together without fermions, stemming from a confining dark $SU(N)$ gauge theory extrapolated from Quantum Chromodynamics (QCD). These glueballs in the dark sector are hypothesized to interact with standard model particle mainly through gravitational forces and may contribute significantly to the dark matter content~\cite{Carlson:1992fn,Boddy:2014yra,Kribs:2016cew,Dienes:2016vei,Carenza:2022pjd,Carenza:2023eua}.

Glueballs within QCD  are bound states of gluons, characterized by the absence of quark constituents. These entities emerge from non-Abelian gauge theories, notably $SU(N)$ lattice gauge theories, where  non-perturbative effects support the formation of such composite particles~\cite{Soni:2016gzf,Soni:2016yes,Acharya:2017szw}. This feature surprisingly bridges the gap between the domain of strong interactions and cosmological phenomena, suggesting a novel view where the dynamics of the strong force support the mysterious nature of dark matter. The similarities underscore the appealing hypothesis that glueballs may indeed constitute the dark matter component in these scenarios~\cite{Carenza:2022pjd}. Theoretical studies and simulations within this framework have been important in substantiating the glueball dark matter hypothesis, offering a robust statement that integrates the microcosmic interactions of particle physics with the large-scale structures and dynamics observed in the universe.

In nonperturbative studies of QCD glueball, lattice QCD calculations play a crucial role~\cite{Morningstar:1999rf,APE:1987ehd,Chen:2005mg,Gui:2012gx,Athenodorou:2020ani,Sun:2017ipk,Gui:2019dtm,Zou:2024ksc}. This non-perturbative method offers detailed insights into key properties of glueballs, including the spectrum and  decay properties. Consequently, it reveals their potential significance across a spectrum of physical phenomena.  For the lattice investigation of self-interacting dark matter, the ratio of  two-body scattering cross section to the mass of dark matter, denoted as $\sigma/m$ is of significant interest. It is not just a crucial quantity for experimental physicists seeking to constrain the characteristics of dark matter \cite{Yoshida:2000uw, Miralda-Escude:2000tvu, Sand:2003ng} but also holds key importance in understanding cosmic phenomena, as highlighted in previous research \cite{Spergel:1999mh}. Recent advances in lattice calculations~\cite{Yamanaka:2019yek,Yamanaka:2019aeq} have been also crucial in determining  the scattering cross section of dark glueballs within lattice $SU(2)$ Yang-Mills theory. 

It should be noticed that traditional Monte Carlo (MC) methods for lattice simulations face issues such as critical slowing down~\cite{Wolff:1989wq} and topological freezing~\cite{Kanwar:2020xzo}. Due to the chain-like nature of the Markov process, configurations generated by traditional methods are not strictly independent unless  a very long  Markov chain is achieved.  Many  improvements are developed in the past decades,  but another potentially  promising method that was recently inspired from machine learning concepts and techniques  is the flow-based model~\cite{Albergo:2019eim,Boyda:2020hsi,Albergo:2021bna,Abbott:2022zhs} (for a review, see Refs.~\cite{Albergo:2021vyo,Cranmer:2023xbe}).
A flow-based model is a type of neural network that can learn a complicated probability distribution from data, and generate new samples from it, without using Markov chain Monte Carlo methods. Moreover, flow-based model can also learn the potential features and correlations of the data, and provide insights into the structure and dynamics of the system. Flow-based models impose an invertible transformation constructed by neural networks transforming a simple distribution to a complex one to perform the maximum likelihood estimation. 

In the remnant of this work, we will use the traditional Monte Carlo method  and  machine learning approach, in a comparative manner,  to generate lattice configurations for the $SU(2)$ gauge theory with zero fermion flavors ($N_f=0$) and several coupling constants $\beta$. Subsequently, we compare the results for the dark glueball obtained from these distinct simulations. Furthermore, we calculate the interaction potential for dark glueball and determine the interaction coupling constants in effective Lagrangian. Finally, we obtain the relation between the dark glueball mass and scattering cross section. By confronting theoretical results with  experimental data, we extract the lower bound on the mass of scalar glueball as candidates of dark matter.  Some details are collected in the appendix. 

\section{Methodology}
In this section, we will  exhibit the theoretical basis of our lattice simulations. This comprises the application of the $SU(N)$ gauge theory action within both the Monte Carlo method and machine learning approach. We elaborate on the specific machine learning methodology, especially the flow-based model. Additionally, this section also includes the strategy  to  investigate the  dark glueball, displaying how to determine its mass, coupling constants of effective Lagrangian, and scattering cross section.

\subsection{$SU(N)$ lattice gauge theory}

The $SU(N)$ gauge theory was established for describing interactions of non-Abelian gauge boson, with $SU(3)$ gauge theory being crucial for modeling strong interactions. Explicitly one can discretize  the spacetime  to simulate QCD in continuum with   gauge links $U_{\mu}(n)$ in path integral approach~\cite{Wilson:1974sk}. Standard simulations of $SU(N)$ gauge theory make use of plaquettes $U_{\mu\nu}(n)$, defined as:
\begin{align}
U_{\mu\nu}(n)=U_{\mu}(n)U_{\nu}(n+\hat{\mu})U^{\dagger}_{\mu}(n+\hat{\nu})U^{\dagger}_{\nu}(n),
\end{align}
with $n$ indicating the lattice coordinates, and $\mu,\nu$ representing gauge link directions. The discretized action is constructed as:
\begin{align}
S[U]=\frac{\beta}{N}\sum_{n,\mu<\nu}\mathrm{Re}\;\mathrm{tr}\left[1-U_{\mu\nu}(n)\right],
\end{align}
where $\beta$ is related to the coupling constant:  $\beta=2N/g^2$. $N$ signifies the color degree of freedom within the $SU(N)$ gauge field, which is set to 2 in this study. The action described previously is referred to as the naive action.

To reduce lattice spacing discretization effects, the Lüscher-Weisz (improved) action is frequently employed~\cite{Luscher:1984xn}, in which both plaquettes $U_{\mu\nu}(n)$ and rectangles $R_{\mu\nu}(n)$ are taken into account: 
\begin{align}
R_{\mu\nu}(n)=U_{\mu}(n)U_{\mu}(n+\hat{\mu})U_{\nu}(n+2\hat{\mu})\nonumber\\
U^{\dagger}_{\mu}(n+\hat{\mu}+\hat{\nu})U^{\dagger}_{\mu}(n+\hat{\nu})U^{\dagger}_{\nu}(n). 
\label{eq:rec}
\end{align}
The corresponding action is constructed as: 
\begin{align}
S[U]=\frac{\beta}{2N}\mathrm{Re}&\Big[\sum_{n,\mu<\nu}c_0\mathrm{tr}\left(1-U_{\mu\nu}(n)\right)\nonumber\\
+&\sum_{n,\mu,\nu}c_1\mathrm{tr}\left(1-R_{\mu\nu}(n)\right)\Big],
\end{align}
where the coefficients for plaquettes and rectangles are $c_0=5/3$ and $c_1=-1/12$, respectively. 

Our study aims to apply self-interacting gauge fields to explore dark matter, where we opt for simplicity by setting $N=2$. We plan to employ both naive and improved actions for Monte Carlo simulations, and the naive action in combination with machine learning simulations, as detailed in Section~\ref{sec:Numerical}.

\subsection{Flow-based model}
In the flow-based model~\cite{Albergo:2019eim, Boyda:2020hsi, Albergo:2021bna, Abbott:2022zhs}, the process begins with a basic prior distribution $r(z)$ that is easy to sample. Utilizing this, a collection of samples ${z_i}$ can be produced and subjected to a reversible one-to-one transformation $f(z)$. Consequently, the transformed samples $x_i=f(z_i)$  will inherently conform to a new distribution $q(x)$:
\begin{align}
q(x)=r(z)[J(z)^{-1}]=r(z)\left | {\det}\frac{\partial f(z)  }{\partial z  } \right |^{-1}, z=f^{-1}(x). 
\end{align}

The objective is to align the new distribution $q(x)$ more closely with the target distribution $p(x)$, which takes the form $p(x)=e^{-\beta S(x)}/Z$ for lattice gauge theory. To assess this alignment,  one can make use of the Kullback-Leibler   (KL) divergence~\cite{Kullback:1951zyt}. The KL divergence between two distributions $q(x)$ and $p(x)$, denoted as $D_{KL}(q|p)$,  is defined as
\begin{align}
    D_{\text{KL}}(q | p) = \int dx\;q(x) \log \left( \frac{q(x)}{p(x)} \right) \geq 0. 
\end{align}
A low KL divergence suggests a close similarity between the two distributions, and $D_{KL}(q|p)=0$ only when $q(x)=p(x)$. In this context, one can construct the transformation as a sequence of neural networks and employ the KL divergence between the resultant distribution $q(x)$ and the target distribution $p(x)$ as a loss function for network training. Throughout the training phase, the KL divergence can be approximated  using the sample mean value: 
\begin{align}
    loss=E_q \left(  \log \frac{q(x)}{p(x)}\right)
    \approx \frac{1}{N} \sum_{i=1}^N \log \frac{q(x_i)}{p(x_i)}. 
\end{align}

Within the flow-based model, the design of the invertible transformation $f(z)$ demands careful consideration due to the Jacobian determinant of $f(z)$ influencing the resulting distribution $q(x)$. To optimize efficiency, the transformation is structured to yield an upper or lower triangular matrix for the Jacobian. 
More explicitly,  one can firstly  choose certain components of $z$ to transform, such as $z^1, z^2...z^k$  (where upper indices denote different components of the variable), and keep the remaining components $z^{k+1},z^{k+2}...z^n$ to be frozen, i.e. $f(z^{j})=z^{j} , k+1 \leq j \leq n$. The overall transformation, which is composed of many single transformations with different frozen freedoms, can transform the samples completely,  such that   every component of $z$ is transformed at least once. Next, one can  define a set of invertible one-to-one mappings $F(z; \alpha_i)$  for $1 \leq l \leq k$ 
with the $\alpha_i$s being parameters. For example, if $z^l \in \mathbb{R}$, one can set the one-to-one maps as $F(z^l; \alpha_i)=\alpha_1 z^l + \alpha_2$ with $\alpha_1 > 0$.  Subsequently, one can construct neural networks that take the fixed components as inputs and output the parameters of the mappings, i.e., the networks map the fixed components to the parameters:
\begin{align}
\{\alpha_i\} = F_{NN}(z^{k+1},z^{k+2}...z^n; \omega_\lambda). 
\end{align}
Here $F_{NN}$  denotes the neural network transformation, which is typically non-invertible or difficult to invert, while the transformation applied to the samples remains invertible. The neural network parameters, denoted as $\omega_\lambda$, need to be optimized. Incorporating fixed components as inputs to the neural networks enhances the overall transformation’s expressiveness and fosters correlations among the various components.

Under the above transformation, the Jacobian is structured as follows:
\begin{align}
\mathbf{J}= \begin{bmatrix}
\frac{\partial x^i}{\partial z^j} \delta_{ij} & \frac{\partial x^l}{\partial z^j} \\
\mathbf{0} & \mathbf{I}_{n-k}
\end{bmatrix}, 1 \leq i,j \leq k, k+1 \leq l \leq n. 
\end{align}
This results in an upper triangular matrix whose determinant can be efficiently computed.

By following these steps, one can illustrate the implementation of the flow-based model in 4-dimensional $SU(2)$ lattice gauge theory. For further insights, refer to ~\cite{arXiv:2305.02402}. The prior distribution is established as a uniform distribution with respect to the Haar measure of the $SU(2)$ group due to its numerical sampling convenience.

To start, the fixed components of the gauge fields should be chosen.  For a 4-dimensional lattice with coordinates $n_i, i=0,1,2,3, n_i \in \mathbb{Z}$, a translational phase is defined based on a translation vector $\mathbf{t}$, where  $t_i \in \mathbb{Z}$: 
\begin{align}
\phi(n) = \sum_i n_i t_i \mod 4. 
\end{align}
This method divides the entire lattice into four segments based on their respective phase values. During a single transformation, one direction of the link variables within a segment is designated as the actively transformed component, while the others are kept fixed. By combining a minimum of 16 individual transformations, a comprehensive transformation is accomplished. In our numerical experiments, we set the translation vector to $(0,1,2,1)$ and designate $U_0(n)$
  as the active transformed link direction.

Secondly, one-to-one mappings are designed. In gauge theory, one needs to preserve the gauge invariance of the theory. Therefore, as link variables are not gauge covariant, they are not directly transformed. Instead, the plaquettes containing the actively transformed links are the ones that undergo transformation. Describing the one-to-one mapping  on plaquettes or in $SU(2)$ group space directly in numerical terms is challenging. However, by considering the theory’s requirement for gauge covariance, we can confine the mapping to be $f(XPX^{-1}) = Xf(P)X^{-1}$, where $P$ and $X$ belong to $SU(2)$. It can be proved that, given the diagonalization of $P$, $P=SUS^{-1}$ the requirement is equivalent to 
\begin{align}
f(P) = Sf(U)S^{-1}, U = \text{diag} (e^{i\theta_1},e^{i\theta_2}).
\end{align}
Namely  only the eigenvalues need to be transformed. The two eigenvalues are not independent of each other due to the constraint $\det P=1$, which leads to $\theta_1+\theta_2=0$. One can design diagonalization procedure to make $\theta_1 \geq 0,  \theta_2 \leq 0$, then
\begin{align}
P=S \begin{bmatrix}
e^{i\theta} & 0 \\
0 & e^{-i\theta}
 \end{bmatrix} S^{-1},  0 \leq \theta \leq \pi. 
\end{align}
The one-to-one mapping now can be simplified as
\begin{align}
f(P)=S \begin{bmatrix}
e^{if(\theta)} & 0 \\
0 & e^{-if(\theta)}
 \end{bmatrix} S^{-1},  0 \leq f(\theta) \leq \pi. 
\end{align}
The requirement now reduce to that $f(\theta)$ should be a one-to-one map on interval $[0, \pi ]$. In our numerical simulations, we opt for the rational-quadratic neural spline flow method with 5 nodes as the one-to-one mapping on the interval $[0, \pi]$ due to its high flexibility. This choice is based on the work~\cite{Durkan:2019nsq}.

In constructing the neural network, maintaining gauge covariance necessitates that inputs consist solely of plaquettes containing frozen link variables. To ensure the translation invariance and cyclic boundary conditions inherent in lattice theory, the optimal architecture involves using convolutional neural networks (CNNs) with cyclic padding. In our simulations, a 4-dimensional CNN with a kernel size of 3 and 48 hidden channels is utilized, employing LeakyReLU as the activation function and the Adam optimizer for training. A mask, shaped to match the lattice plaquettes, is applied to select the frozen plaquettes before they are inputted.


It is necessary to point out that although the flow-based model has shown great success, it encountered the mode collapse problem when $\beta$ was increased~\cite{Nicoli:2023qsl}. To address this problem, a quenched method was proposed in a study in Ref.~\cite{Abbott:2023thq}.  
The quenched method involves first training a model to generate samples that adhere to the distribution $e^{-\beta_1 S(x)}/Z_1$, where $\beta_1$
is small. Subsequently, this distribution with a small $\beta_1$ is utilized as the prior distribution to train a second model that generates samples following $e^{-\beta_2 S(x)}/Z_2$ with a larger $\beta_2$. This process is iterated until the value of $\beta$ reaches the desired level. In this work we will  start with  $\beta=0$ and   incrementally increase $\beta$ by 0.1 or 0.2 every 10 or 20 rounds of training until $\beta$ reaches the target value. Through this way the total training time is reduced compared to the original method, and an advantage is that  only one model needs to be saved. The overall training time increases approximately linearly with $\beta$ and the lattice volume (the number of training parameters increases approximately linearly with the lattice volume), while the time required for generating configurations is only dependent on the lattice volume.  At this point, it is important to note that the flow-based model is inefficient in generating configurations with a very large volume. For a comparative analysis in this work, we set $\beta$ up to 2.2 and the lattice volume up to $10^4$.

\subsection{Dark glueballs and scattering}
\subsubsection{Dark Glueball Spectrum}

In lattice simulations, the operator for dark glueball with $J^{PC}=0^{++}$ is given:
\begin{align}
\hat{O}(t,\vec{x})=\mathrm{Re} \;\;\mathrm{tr}\sum_{\mu<\nu}U_{\mu\nu}(t,\vec{x}). 
\label{eq:operator}
\end{align}
To account for the vacuum state sharing the glueball's quantum numbers  $J^{PC}=0^{++}$, the vacuum expectation value is subtracted, reformulating the operator as $\hat{O}^{\prime}(t,\vec{x})=\hat{O}(t,\vec{x})-\langle\hat{O}(t,\vec{x})\rangle$. We can use two-point correlation functions to determine the mass of dark glueball,
\begin{align}
C_2(t)=\sum_{\vec{x}, \vec{y}}\langle0|\hat{O}(t,\vec{x})\hat{O}(0,\vec{y})|0\rangle.
\label{eq:2pt}
\end{align}
When the time slice $t$ is large enough, the two-point correlation function is reduced to
\begin{align}
C_2(t) \sim C_0e^{-E_0t}.
\label{eq:2pt_reduced}
\end{align}
The effective energy $E_0$  represents the energy of the lowest dark glueball state (ground state) and when the momentum is zero, it is reduced to the mass $m_g$.  

In contrast to the naive action, the operator formulation for the dark glueball in the context of the improved action incorporates an additional rectangular term, as defined in Eq.(\ref{eq:rec}). Thus the operator becomes:
\begin{align}
\hat{O}_{\mathrm{imp}}(t,\vec{x}) = \mathrm{Re}\mathrm{tr}\sum_{\mu<\nu}\left[U_{\mu\nu}(t,\vec{x}) + \frac{1}{20}R_{\mu\nu}(t,\vec{x})\right].
\label{eq:operator}
\end{align}
For subsequent analyses, we employ this refined operator representation for the dark glueball within the framework of configurations derived from the improved action.

\subsubsection{The two body scattering cross section}

In the following, we will determine the two-body dark  glueball scattering cross section in various ways. 

The cross section of dark glueballs scattering can be inferred from the interaction glueball potential in a non-relativistic limit. 
In an approach one can employ the Bonn approximation to determine the cross section from the interaction glueball potential. This potential is modeled in this work,  for instance,  as  a spherically symmetric Yukawa and Gaussian potential:
\begin{gather}
V_{\mathrm{Yukawa}}(r)=Y\frac{e^{-m_gr}}{4\pi r},\label{eq:yukawa}\\
V_{\mathrm{Gaussian}}(r)=Z^{(1)}e^{-\frac{(m_gr)^2}{8}}+Z^{(2)}e^{-\frac{(m_gr)^2}{2}}.\label{eq:gaussian}
\end{gather}
where $m_g$ represents the mass of the dark glueball, and $Y, Z^{(1)}, Z^{(2)}$ are parameters. For both Yukawa and Gaussian potentials, one can consider   the  differential cross section in nonreltivistic limit with $k\to0$ which is derived as follows:
\begin{align}
\frac{d\sigma}{d\Omega}=\frac{m^2_g}{|\vec{K}|^2}\left|\int_0^{\infty}rV(r)\sin(|\vec{K}|r)dr\right|^2. 
\label{eq:bonn_app}
\end{align}
Here $\vec{K}=\vec{p}_3-\vec{p}_1$ represents momentum transfer, and $|\vec{K}|=2k\sin\frac{\theta}{2}$ with $k$ and $\theta$ being the magnitude and scattering angle of the relative momentum $\vec{k}$ between two interacting particles respectively. $V(r)$ denotes either the Yukawa or the Gaussian potential.

Alternatively, using an effective theory for scalar particles, and considering the $\phi^3$ and $\phi^4$ interaction, one can construct the following Lagrangian for dark glueballs as:
\begin{align}
\mathcal{L}=\frac{1}{2}(\partial_{\mu}\phi)^2-\frac{1}{2}m_g^2\phi^2-\frac{\lambda_3}{3!}\phi^3-\frac{\lambda_4}{4!}\phi^4.
\end{align}

\begin{figure}
\centering
\includegraphics[scale=0.15]{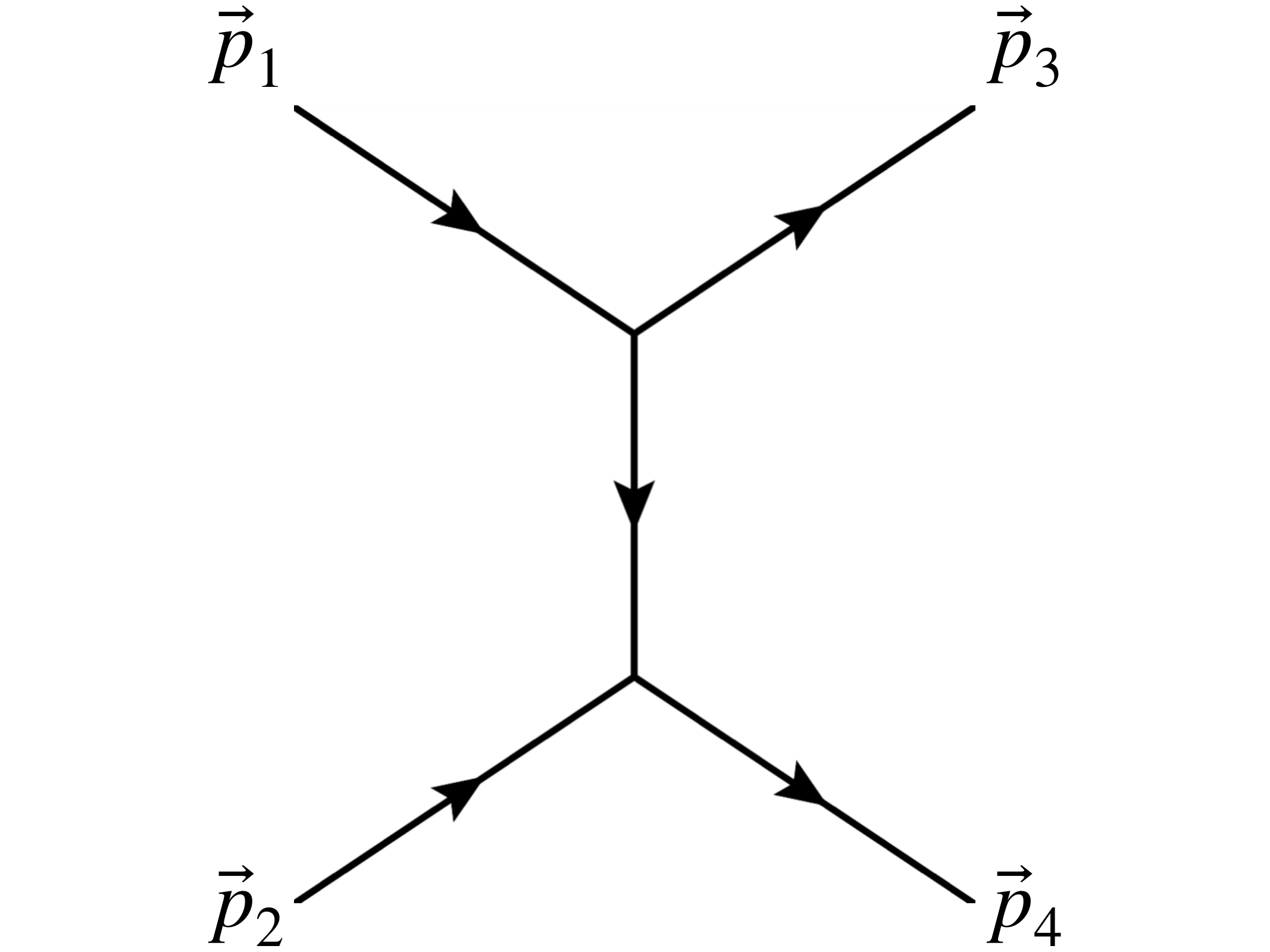}
\caption{The Feynman diagram of $t$ channel for elastic scattering of scalar field particles.}
\label{fig:scalar_fm}
\end{figure}

\noindent Corresponding to the contributions in the potential, the $\phi^4$ interaction term gives a delta function, while it is difficult to be obtained numerically. Therefore, only $\phi^3$ term is considered for the scattering. Perturbative theory yields the differential cross section for two-body elastic scattering $\phi(\vec{p}_1)\phi(\vec{p}_2)\to\phi(\vec{p}_3)\phi(\vec{p}_4)$. The Feynman diagram in Fig. \ref{fig:scalar_fm} provides the $i\mathcal{M}$ matrix element:
\begin{align}
i\mathcal{M}=(-i\lambda_3)^2\frac{i}{-(\vec{p}_3-\vec{p}_1)^2-m_g^2}=\frac{i\lambda_3^2}{|\vec{K}|^2+m_g^2}.
\label{eq:dsigma_phi3}
\end{align}
Setting $|\vec{K}|=0$, we obtain the differential cross section:
\begin{align}
\frac{d\sigma}{d\Omega}=\frac{|\mathcal{M}|^2}{64\pi^2(2m_g)^2}=\frac{\lambda_3^4}{256\pi^2m_g^6}.
\label{eq:dsigma_phi3_k0}
\end{align}
The $\phi^3$ interaction generates Yukawa potential in the coordinate space in non-relativistic limit. If $V_{\mathrm{Yukawa}}(r)$ is inserted to Eq.(\ref{eq:bonn_app}), the differential cross section is expressed as:
\begin{align}
\frac{d\sigma}{d\Omega}\Bigr\|_{\mathrm{Yukawa}}=\frac{Y^2}{16\pi^2m_g^2}.
\label{eq:dsigma_yukawa_k0}
\end{align}
Comparing Eq.(\ref{eq:dsigma_phi3_k0}) and Eq.(\ref{eq:dsigma_yukawa_k0}), one can find that  the value of $\lambda_3$ is determined by the coefficient $Y$:
\begin{align}
\lambda_3=2m_g\sqrt{|Y|}.
\label{eq:lambda3_Y}
\end{align}

The $s$-wave scattering cross section  can also be directly  extracted from the wave function. This method is grounded in the radial Schrödinger equation, in which the wave function asymptotically approaches the asymptotic form as $r \to \infty$:
\begin{align}
\Psi(r)\xrightarrow{r\to\infty}\frac{A_{l}}{r}\sin\left[kr+\delta_l(k)\right].
\end{align}
By solving the radial Schrödinger equation, we can extract the wave function $\Psi(r)$ and determine the phase shift $\delta_l(k)$. This enables the calculation of both the differential and total cross sections for the $s$-wave scattering in the $k\to0$ limit:
\begin{align}
\frac{d\sigma_s}{d\Omega}&=\lim_{k\to0}\frac{1}{k^2}\sin^2\left[\delta(k)\right], \\
\sigma&\sim\sigma_s=4\pi\frac{d\sigma_s}{d\Omega}.
\end{align}

\subsubsection{Nambu-Bethe-Salpeter wave function}
To calculate the coefficient $Y$ in the Yukawa potential or $Z^{(1)}$ and $Z^{(2)}$
in the Gaussian potential, we make use of the Nambu-Bethe-Salpeter (NBS) wave function derived from lattice calculations. The NBS wave function was initially developed for hadrons \cite{Aoki:2013cra,Ishii:2012ssm}, which is also suitable for analyzing glueballs. It is defined as:
\begin{align}
\Psi_g(\vec{r})=\sum_{\vec{x}}\left\langle0\left|\hat{O}'(0,\vec{x}+\vec{r})\hat{O}'(0,\vec{x})\right|GG\right\rangle,
\end{align}
where $\hat{O}'$ is a glueball operator, and $|GG\rangle$ represents a hadron state comprising two glueballs.

The NBS wave function related to the Schrödinger equation with the interaction dark glueball potential. The spatial Schrödinger equation with a non-local potential is expressed as:
\begin{align}
\frac{1}{m_g}\nabla^2\Psi_g(\vec{r})=\int d^3r^{\prime}U(\vec{r}^{\prime},\vec{r})\Psi_g(\vec{r}^{\prime}),
\end{align}
where $U(\vec{r}^{\prime},\vec{r})$ approximates a local, central potential $U(\vec{r}^{\prime},\vec{r})=V(r)\delta^3(\vec{r}^{\prime}-\vec{r})$. The relation between $V(r)$ and $\Psi_g(\vec{r})$ is:
\begin{align}
V(r)=\frac{1}{m_g}\frac{\nabla^2\Psi_g(\vec{r})}{\Psi_g(\vec{r})}.
\label{eq:pontential_NBS}
\end{align}
Replacing $V(r)$ with Eq.(\ref{eq:yukawa}) or Eq.(\ref{eq:gaussian}) allows the determination of the coefficient $Y$ or $Z^{(1)}$, $Z^{(2)}$ from $\Psi_g(\vec{r})$.

In lattice simulations, correlation functions to extract the NBS wave function are constructed as:
\begin{align}
C_g(t,\vec{r})=\sum_{\vec{x}}\left\langle0\left|\hat{O}'(t,\vec{x}+\vec{r})\hat{O}'(t,\vec{x})\hat{S}(0,0)\right|0\right\rangle,
\label{eq:correlation}
\end{align}
with $\hat{S}$ as the glueball operator  source. Then, the wave function is given as
\begin{align}
\Psi_g(\vec{r})=\lim_{t\to\infty}\frac{C_g(t,\vec{r})}{C_g(t,\vec{r}=0)}.
\end{align}

In principle a two-body state could serve as the source $\hat{S}$. However for the dark glueball with $J^{PC}=0^{++}$, $|GG\rangle$ and $|G\rangle$ share the same quantum numbers. Additionally, one cannot split one-body source into spatially separated correlators, unlike two- and three-body sources. Hence, we use  $\hat{S}=\hat{O}'$.

\section{Numerical simulations and results}
\label{sec:Numerical}

In this section, we give our numerical results obtained from lattice simulations. This includes an in-depth overview of the configuration generation process, the results for dark glueball mass, as well as the determination of the coupling coefficient for the effective Lagrangian and the dark glueball scattering cross section.

\subsection{Lattice setup}

When generating lattice configurations,
we have varied the coupling constant $\beta$ in the set $\{2.0, 2.2, 2.4, 2.6\}$ for naive actions in $SU(2)$ simulations.  In contrast, for the improved actions, we specifically employed $\beta=3.5$ for comparative purposes.

\begin{table}[htbp]
\centering
\caption{Setup for $SU(2)$ Lattice simulations: The initial configurations are random $SU(2)$ matrices, followed by warm-up process with heatbath updates, and then production process employing combinations of one heatbath update and multiple overrelaxation updates.} 
\setlength\tabcolsep{0pt}
\begin{tabular*}{0.48\textwidth}{@{\extracolsep{\fill}} ccccccc}
\hline \hline
action    & $\beta$ & Volume         & heat+over & Warm up & Conf \\ \hline
naive     & 2.0     & $10^4$         & 1+1       & 500            & $4608$ \\ 
naive     & 2.2     & $10^4$         & 1+2       & 700            & $4608$ \\ 
naive(ML) & 2.2     & $10^4$         & /         & /              & $4608$ \\ 
naive     & 2.4     & $16^4$         & 1+3       & 700            & $2304$ \\ 
naive     & 2.6     & $24^4$         & 1+5       & 1000           & $1536$ \\ 
improved  & 3.5     & $16^4$         & 1+5       & 1000           & $1536$ \\ \hline
\end{tabular*}
\label{tab:setup}
\end{table}

Table. \ref{tab:setup} delineates the setup of configurations employed in our study. During the generation process, we have used  both heatbath and overelaxation updates for the $SU(2)$ gauge field. To enhance signal clarity, we implemented APE (Array Processor with Emulator) smearing for the gauge links as described in \cite{APE:1987ehd},
\begin{align}
&V_{\mu}(n) = U_{\mu}(n) + \alpha \sum_{\pm\nu \neq \mu} U_{\nu}(n) U_{\mu}(n+\hat{\nu}) U^{\dagger}_{\nu}(n+\hat{\mu}),\\
&U^{\mathrm{smear}}_{\mu}(n) = V_{\mu}(n)/\sqrt{\det{V_{\mu}(n)}}.
\end{align}
In our simulations, we select a smearing parameter of $\alpha = 0.1$, and a total of 20 APE smearing iterations are uniformly applied across all configurations. Employing the established relation between the string tension $\sigma$ and the scale parameter $\Lambda$ as documented in \cite{Allton:2008ty}, we determine the lattice spacing in terms of $\Lambda^{-1}$ for each simulation setting \cite{Bali:2000gf}. Detailed information is given in Appendix.\ref{sec:wiloop}, and the results are listed in Table \ref{tab:lattice_spacing}.

\begin{table}[htbp]
\centering
\caption{The lattice spacings in string tension $\sigma$ unit and $SU(2)$ scale $\Lambda$ for configurations of several $\beta$ based on naive and improved action.}
\setlength\tabcolsep{0pt}
\begin{tabular*}{0.48\textwidth}{@{\extracolsep{\fill}} cccc}
\hline \hline
action & $\beta$ & $a\sqrt{\sigma}$ & $a[\Lambda^{-1}]$ \\ \hline
naive    & 2.0     & 0.6365(66)     & 0.3727(39)        \\ 
naive    & 2.2     & 0.4857(30)     & 0.2844(17)        \\ 
naive    & 2.4     & 0.2868(24)     & 0.1621(14)        \\ 
naive    & 2.6     & 0.1669(96)     & 0.0977(56)        \\ 
improved & 3.5     & 0.3246(17)     & 0.1900(10)        \\ \hline
\end{tabular*}
\label{tab:lattice_spacing}
\end{table}

\begin{figure}
\centering
\includegraphics[scale=0.7]{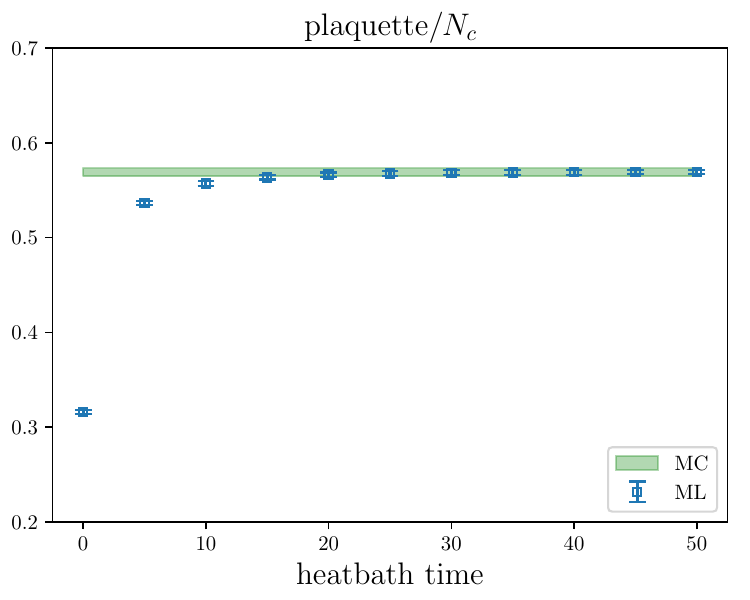}
\caption{The comparison of plaquette based on configurations from Monte Carlo method and machine learning refined  with the heatbath method.}
\label{fig:plaq}
\end{figure}

\begin{figure}
\centering
\includegraphics[scale=0.7]{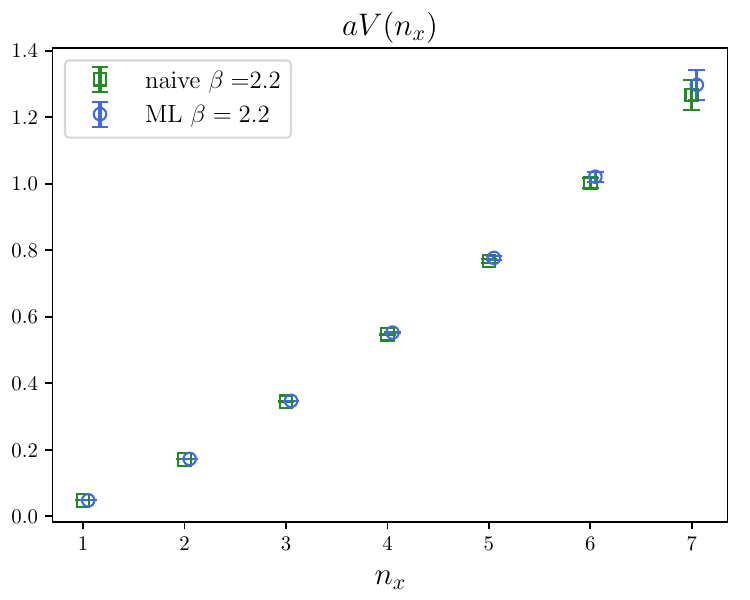}
\caption{The comparison of static potential from MC and ML with 40 times heatbath updates.}
\label{fig:aV}
\end{figure}

In contrast, for the machine learning approach, we restrict our simulations to the naive action at $\beta = 2.2$ with a lattice volume of $10^4$. While machine learning yields configurations that closely resemble $SU(2)$ gauge configurations, they are not precise. Therefore, we use the heatbath method to update the gauge links from machine learning. Fig. \ref{fig:plaq} compares the plaquette values from MC and ML with heatbath updates. This comparison reveals that MC configurations align with ML ones after 25 times heatbath updates. Eventually, we performed 40 times. Furthermore, we have compared the static potential $aV(n_x)$, derived from large Wilson loops as shown in Eq.(\ref{eq:V_WLP}), which is illustrated in Fig. \ref{fig:aV}. In this figure, the static potential displays a positive value of $B$ for the $B/r$ term, which is attributed to APE smearing. Further details are discussed in Appendix \ref{sec:wiloop}. This comparison supports our conclusion that machine learning with heatbath updates can generate reasonable results  as the traditional Monte Carlo method.

\begin{figure}
\centering
\includegraphics[scale=0.7]{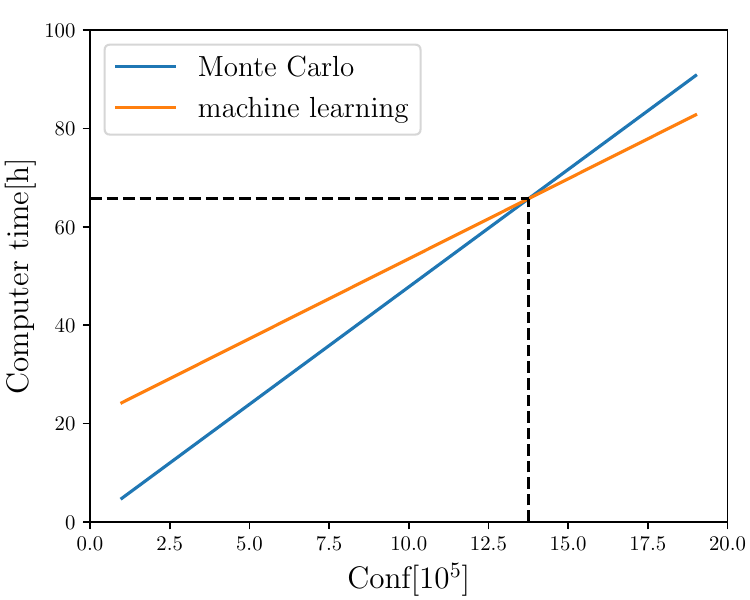}
\caption{Comparison of computer costs from Monte Carlo and machine learning methods.}
\label{fig:cost}
\end{figure}

In this investigation, we also undertake a comparative analysis of the computational time consumption between the Monte Carlo method and machine learning approach.  It should be warned that this comparison is qualitative since for machine learning approach due to the lack of GPU nodes we have used the CPU resources across 6 DCU nodes.   When implementing the Monte Carlo calculation for $10^4$ lattice configurations, the processing time is approximately 0.17 seconds per configuration. This implies that the initial warm-up process requires around 120 seconds, and the total time for generation, including both warm-up and production, is calculated as (Conf$\times$0.17+120) seconds. In contrast, the machine learning method involves distinct training and production phases, consuming 21 hours for training and Conf$\times$0.12 seconds for production. Fig. \ref{fig:cost} illustrates a cost comparison between the Monte Carlo and machine learning methods. Based on this analysis, it can be inferred that the Monte Carlo method is more efficient for small-scale computations, whereas machine learning demonstrates greater suitability for extensive-scale simulations. Therefore, configurations derived from machine learning exhibit reduced correlation and demand less computational time in this example. From this perspective, machine learning may emerge as an efficient method for generation.

On the other side, it is necessary to point out that the implementation of machine learning  is still limited. A significant issue is the substantial memory requirement during the training process, which may exceed the capacity of limited-core systems. Additionally, the persistent problem of mode-collapse in machine learning necessitates the continued use of the Monte Carlo method for configuration filtering. Given these considerations, our primary simulations employ the Monte Carlo approach, while machine learning is only utilized for a comparative analysis at this stage.

\subsection{Dark glueball mass}
\begin{widetext}

\begin{figure}
\centering
\includegraphics[scale=0.6]{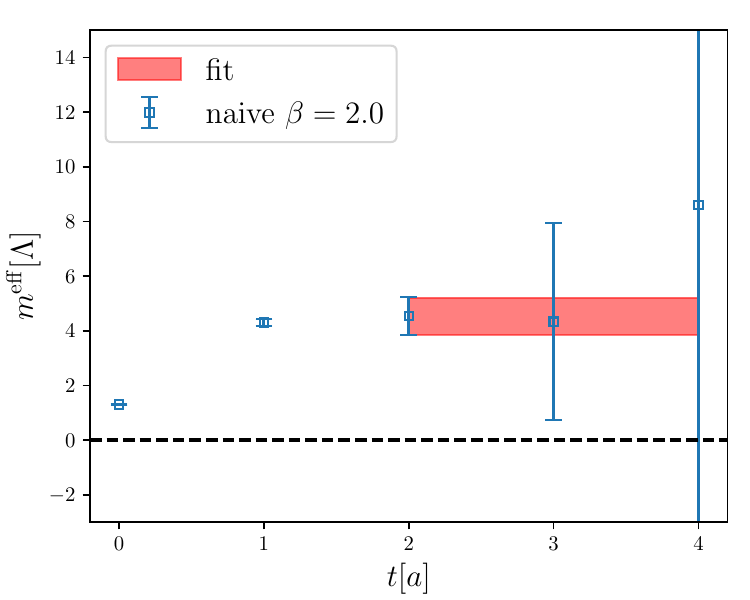}
\includegraphics[scale=0.6]{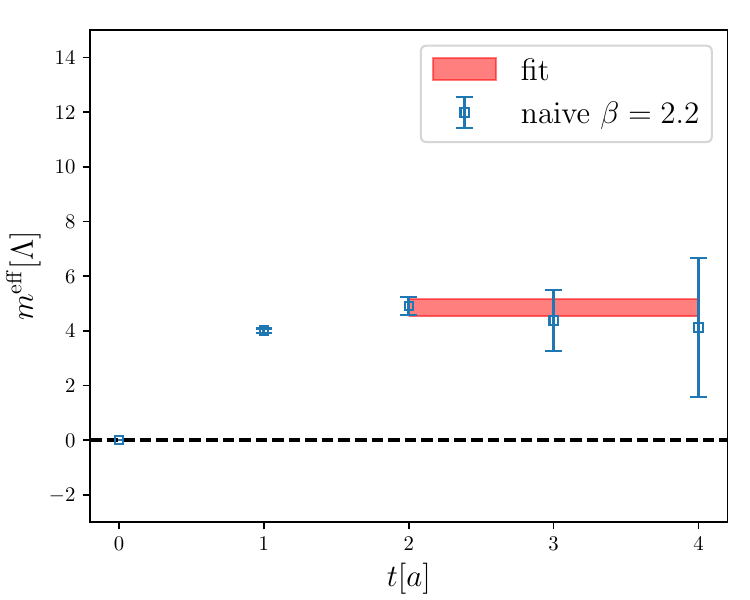}
\includegraphics[scale=0.6]{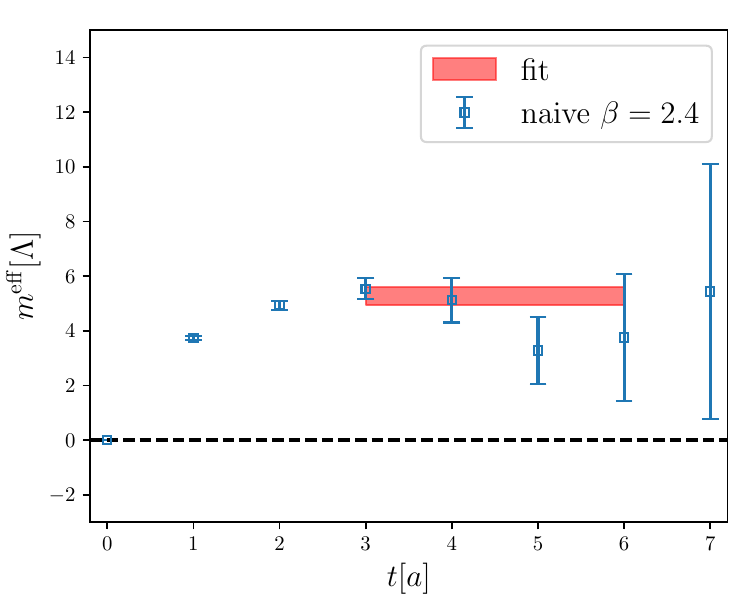}
\includegraphics[scale=0.6]{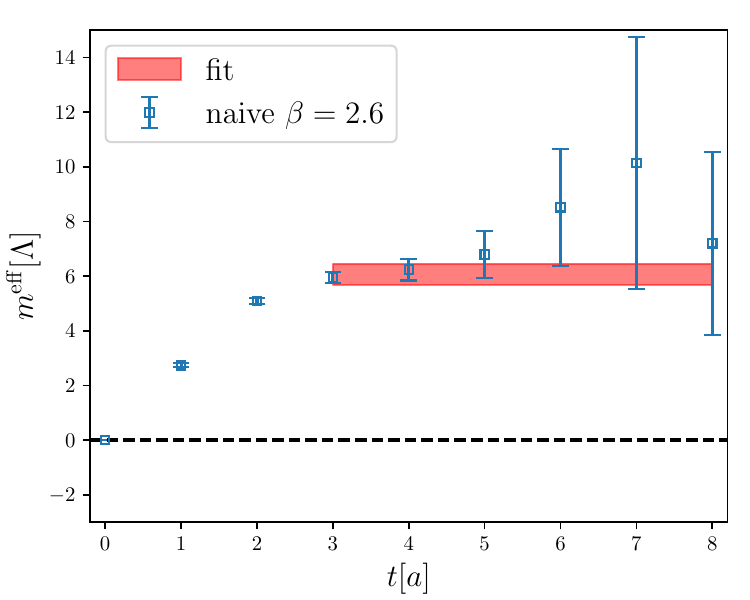}
\includegraphics[scale=0.6]{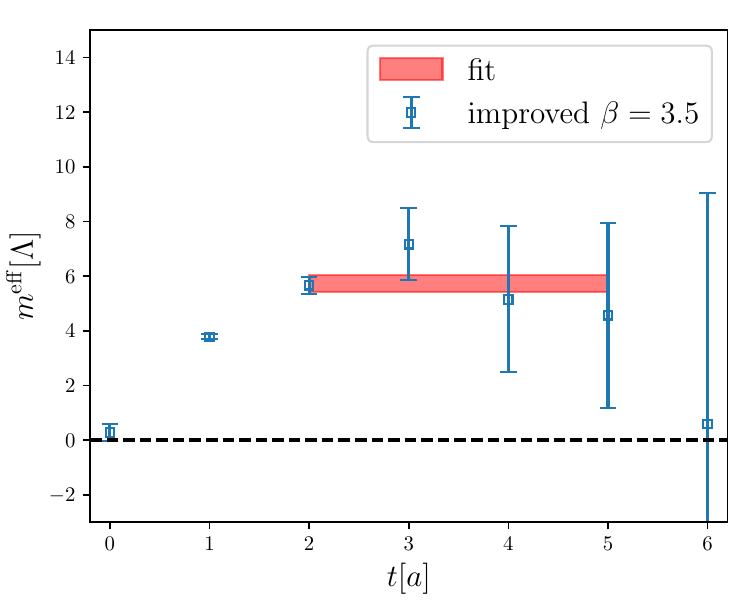}
\includegraphics[scale=0.6]{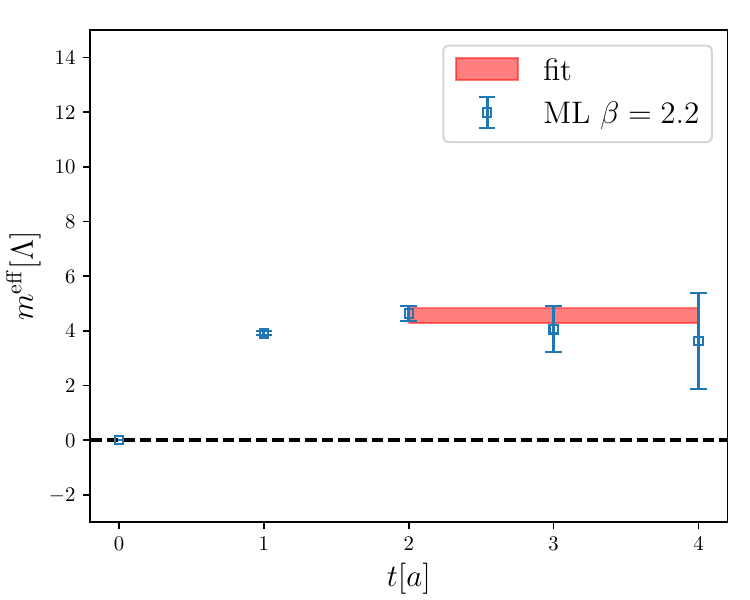}
\caption{Effective mass plots are displayed for all cases. The corresponding fitting results are shown in Table \ref{tab:mass}. The red bands in these plots represent the fitting results, with their lengths corresponding to the fitting ranges.}
\label{fig:mass}
\end{figure}

\end{widetext}

The mass of the dark glueball in the framework of $SU(2)$ gauge theory is ascertainable through the analysis of the two-point correlation function $C_2(t)$, as delineated in Eq.(\ref{eq:2pt}). Utilizing the dark glueball operator provided in Eq.(\ref{eq:operator}), we select $\mu$ and $\nu$ along the spatial directions to characterize the dark glueball. Considering the ground state contribution, $C_2(t)$ exponentially decays with respect to time slice $t$. In light of the periodic boundary conditions applied during the configuration generation, $C_2(t)$ is modeled as:
\begin{align}
C_2(t) \sim C_0e^{-m_gt} + C_0e^{-m_g(N_t-t)}.
\end{align}
To clearly demonstrate the value of the dark glueball mass, we presented the effective mass plots in Fig. \ref{fig:mass}. Here, we calculated the effective mass $m^{\rm eff}$, using the following formula:
\begin{align}
m^{\rm eff}={\rm arccosh } \frac{C_2(t-1)+C_2(t+1)}{2C_2(t)}.
\end{align}
Despite considerable uncertainties  in each case, the trend suggests that as $t$ increases, $m^{\rm eff}$ tends to stabilize roughly towards a plateau. To enhance the stability and clarity of our fitting process, we performed a constant fit for $m^{\rm eff}$. The results are illustrated in Fig. \ref{fig:mass} as red bands, whose lengths represent the ranges of used data in fitting. The results are also listed in Table \ref{tab:mass}, which indicates that as $\beta$ increases, the dark glueball mass exhibits a gradual increase.

\begin{table}[htbp]
\centering
\caption{The results of effective masses of $0^{++}$ glueball are shown in lattice unit and in $\Lambda$.}
\setlength\tabcolsep{0pt}
\begin{tabular*}{0.48\textwidth}{@{\extracolsep{\fill}} cccc}
\hline \hline
action & $\beta$ & $am_g$ & $m_g[\Lambda]$ \\ \hline
naive      & 2.0     & 1.69(25)        & 4.53(68)       \\ 
naive      & 2.2     & 1.373(89)       & 4.83(32)       \\ 
naive(ML)  & 2.2     & 1.293(77)       & 4.56(27)      \\ 
naive      & 2.4     & 0.853(54)       & 5.26(34)       \\ 
naive      & 2.6     & 0.593(17)       & 6.06(39)      \\ 
naive      & continuum limit & /       & 6.35(51)       \\
improved   & 3.5     & 1.087(58)       & 5.72(31)       \\ \hline
\end{tabular*}
\label{tab:mass}
\end{table}

Moreover, we extrapolate the dark glueball mass to the continuum limit by using the equation:
\begin{align}
m^{\rm eff}_g(a) = m^{\rm eff}_g(a\to0)+c_1a+c_2a^2,
\end{align}
and determine the results $m_g(a\to0)=6.35(51)\Lambda$.  Since only the naive action is employed in the extrapolation, we retain the $\mathcal{O}(a)$ term, which should be neglected with improved action. In parallel, our simulations using the improved action suggest an effective glueball mass of $m_g=5.72(31)\Lambda$. Comparing this result with those obtained using the naive action, we find that the improved action configurations at such a large lattice spacing $a=0.1900(10)\Lambda^{-1}$ has smaller discrete effect than naive action. We have put together all these results in Tab. \ref{tab:mass}, which also shows very similar values for the naive action at $\beta=2.2$, achieved through both machine learning and Monte Carlo methods. This comparison confirms that traditional Monte Carlo methods and machine learning techniques are consistent with each other. These results from different methods underline the usefulness of machine learning in particle physics, especially for complicated tasks like calculating the dark glueball mass.

\subsection{Scattering cross section}
\label{sec:scatter}

\begin{figure}
\centering
\includegraphics[scale=0.7]{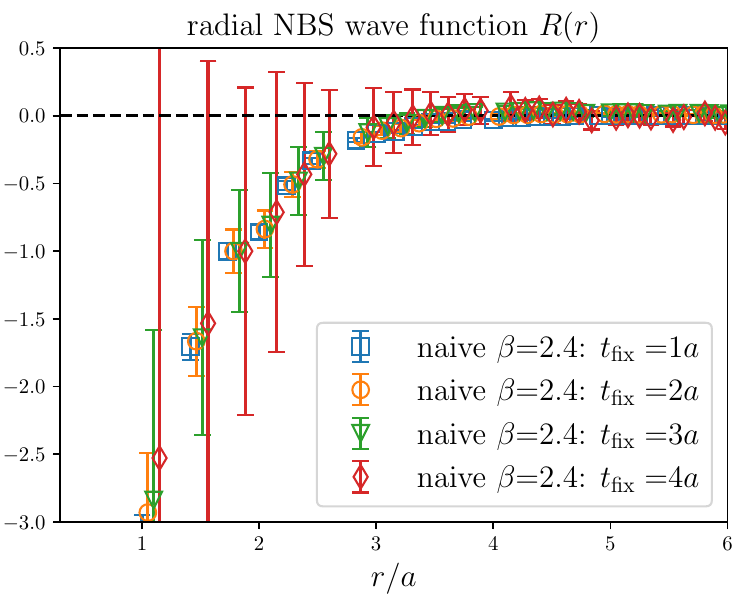}
\caption{Radial NBS wave function $R(r)$ at different time separations.}
\label{fig:NBS_t_dep}
\end{figure}

In this subsection, we address the determination of interaction dark glueball potentials using the NBS wave function~\cite{Aoki:2013cra,Ishii:2012ssm}, as outlined in Eq.(\ref{eq:pontential_NBS}). Tackling the numerical second partial derivative in this context is challenging. In the case that the potential is spherically symmetric, we  can approach it  by solving a radial partial differential equation:
\begin{align}
\frac{d^2R(r)}{dr^2}+\frac{2}{r}\frac{dR(r)}{dr}-m_gV(r)R(r)=0,
\label{eq:ode}
\end{align}
Here, $R(r)$ is the radial part of $\Psi_g(\vec{r})$. The solution's angular component is expressed as spherical harmonics with quantum numbers $l,m$, leading to $\Psi_g(\vec{r})=R(r)Y_{l,m}(\theta,\phi)$. Averaging $\Psi_g(\vec{r})$ over a spherical shell simplifies $Y_{l,m}(\theta,\phi)$ to $1$, effectively reducing the NBS wave function $\Psi_g(\vec{r})$ to $R(r)$ times a constant.

\begin{widetext}

\begin{figure}
\centering
\includegraphics[scale=0.7]{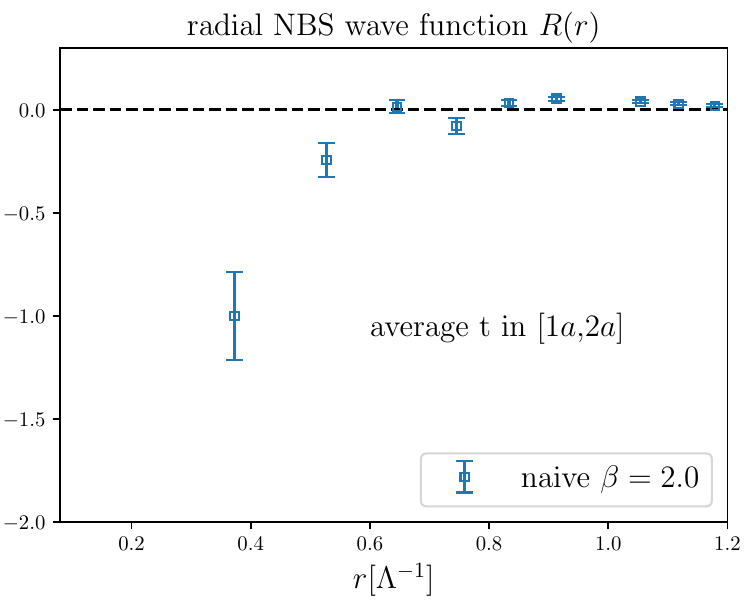}
\includegraphics[scale=0.7]{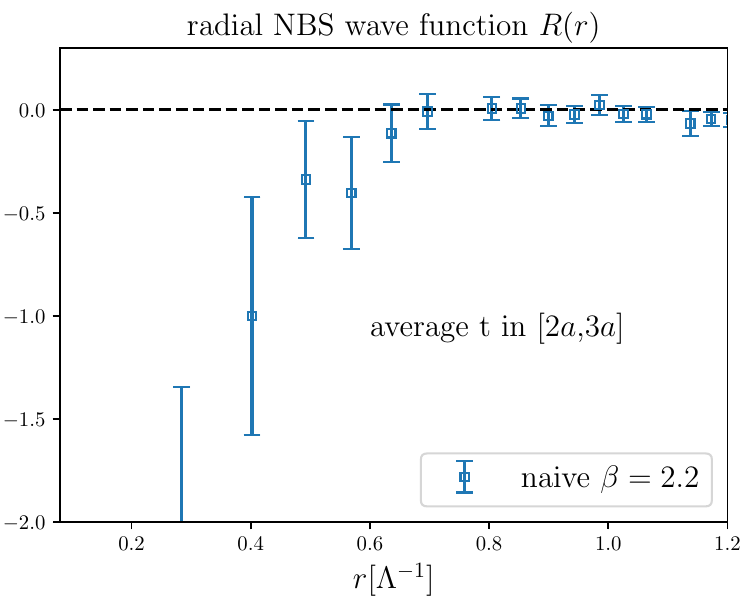}
\includegraphics[scale=0.7]{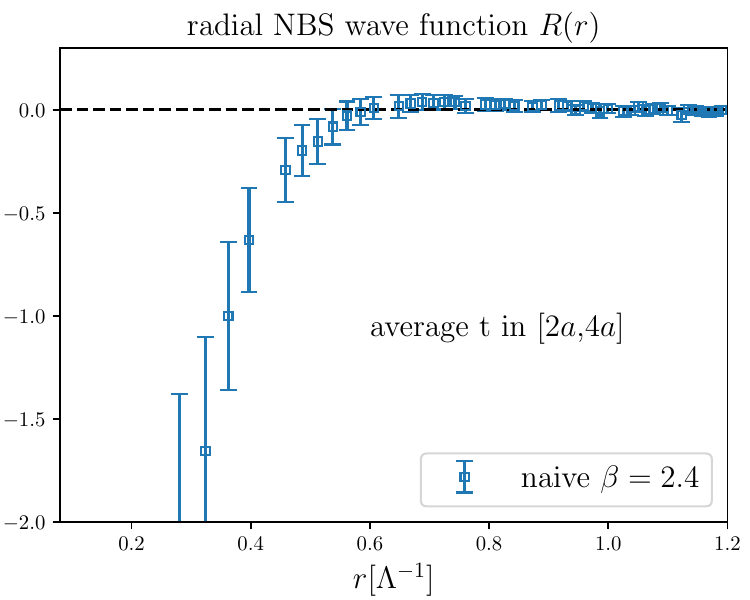}
\includegraphics[scale=0.7]{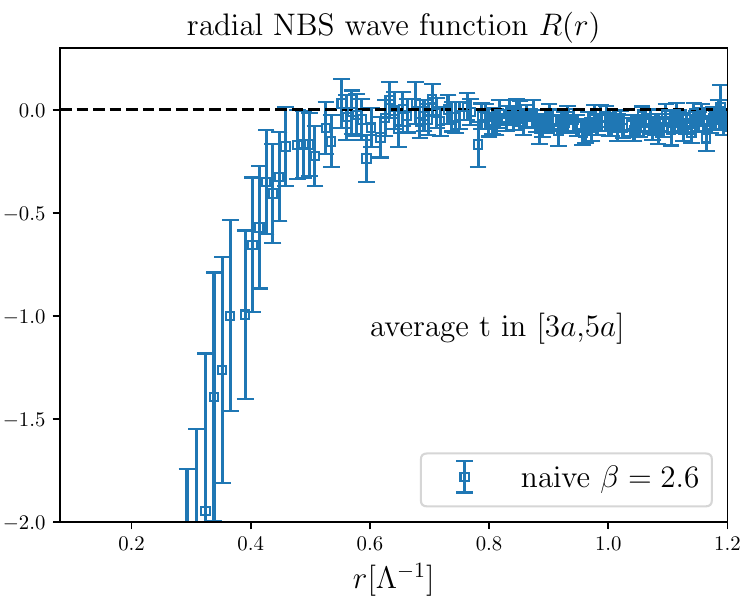}
\includegraphics[scale=0.7]{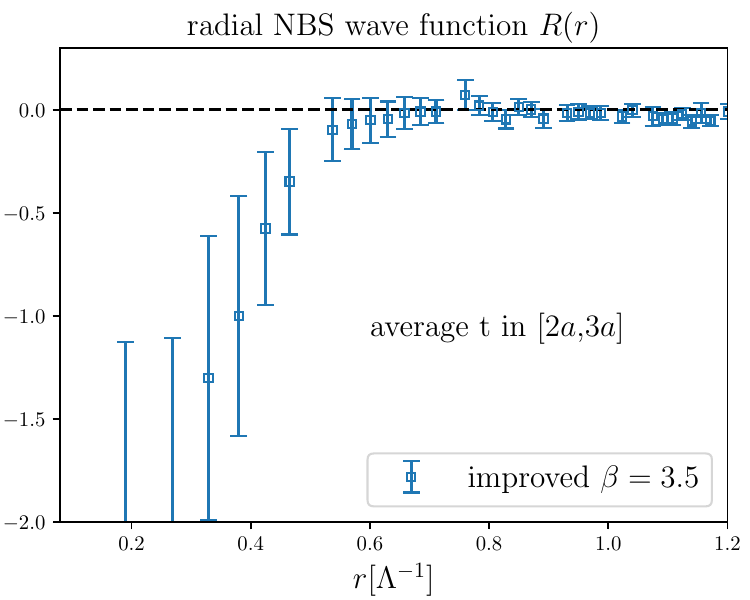}
\includegraphics[scale=0.7]{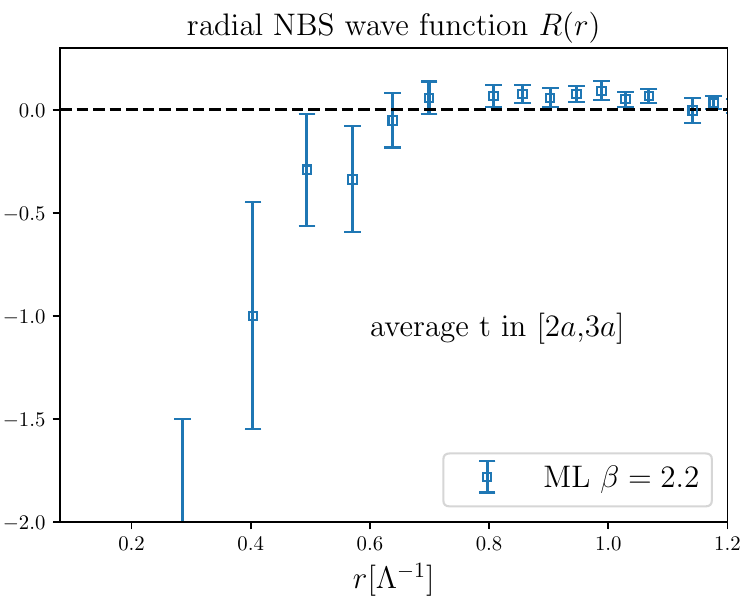}
\caption{This figure shows the radial part of the NBS wave function derived from various lattice setups. We normalize all data at $r=0.3727\Lambda$, and show the chosen $t$ range in the middle of each panel. As noted in the main text, the results labeled `naive' and `improved' are based on configurations with naive and improved gauge action, and `ML' represents results obtained with machine learning methods.}
\label{fig:NBS}
\end{figure}

\end{widetext}

To proceed, we calculate the two-body correlation function as stated in Eq.(\ref{eq:correlation}) and average it over spherical shells, approximating it as $R(r)$. However, it is derived under the condition $t\to\infty$ which cannot be achieved in lattice simulations. Therefore, we examine $R(r)$ at fixed time separation, denoted as $t_{\rm fix}$, which are displayed in Fig. \ref{fig:NBS_t_dep}. We notice that $R(r)$ exhibits modest variation with increasing $t$, prompting us to utilize the averaged results from several $t_{\rm fix}$s. Our findings for $R(r)$ from different lattice setups are presented in Fig. \ref{fig:NBS}.

Considering Yukawa and Gaussian forms of potential energy, we set boundary conditions and adjust coefficients $Y$ or $Z^{(1)}$, $Z^{(2)}$ to find solutions for $R(r)$, noted as $R(r,Y)$ or $R(r,Z^{(1)},Z^{(2)})$ using the fourth-order Runge-Kutta method. We then measure the differences between the $R(r,Y)$ or $R(r,Z^{(1)},Z^{(2)})$ curves and the data of $R(r)$, selecting the curve that most closely matches our data to determine the coefficients.

\begin{figure}
\centering
\includegraphics[scale=0.7]{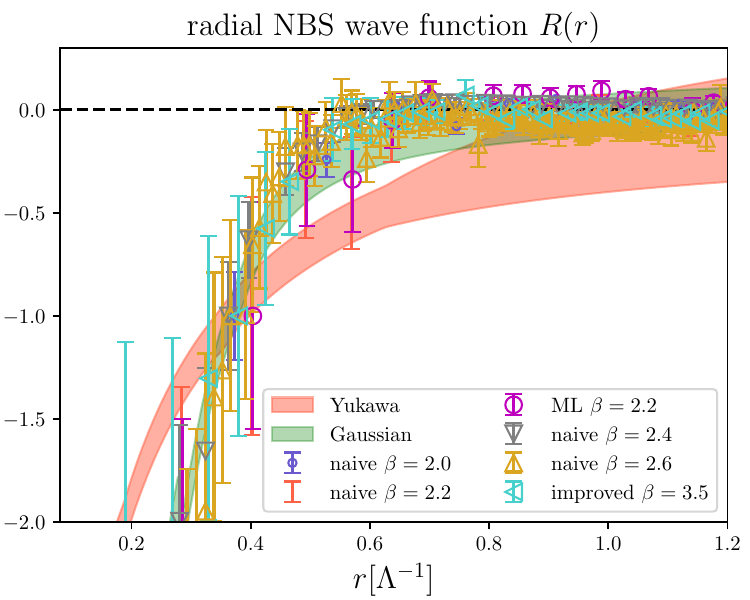}
\caption{The plot shows the results by solving the differential equation with Yukawa and Gaussian potential.}
\label{fig:NBS_final}
\end{figure}

Fig. \ref{fig:NBS_final} compares the solutions with Yukawa and Gaussian potentials to our lattice data. This shows that solving the differential equation Eq.(\ref{eq:ode}) with the Runge-Kutta method aligns well with our simulations, indicating reliable results. We thus obtained the values for $Y$ in the Yukawa potential, and $Z^{(1)}$, $Z^{(2)}$ in the Gaussian potential. In addition, we also investigated the $\chi^2/\text{d.o.f}$ values of both Yukawa and Gaussian fits, finding them to be $2.915$ and $0.764$, respectively, over the fitting range of $r\in[0.4,0.9]\Lambda^{-1}$. 
This indicates that the Yukawa form only partially captures the data.  Consequently, we decided to increase the uncertainty of the fit parameters by a factor of $\sqrt{\chi^2/\text{d.o.f}}$. For clarity, we reformulate these equations as follows:
\begin{align}
&V_{\mathrm{Yukawa}}(r) = -17.6(2.6)\frac{e^{-m_gr}}{4\pi r}, \\
&V_{\mathrm{Gaussian}}(r) \nonumber\\
&= 12.46(0.35)\Lambda e^{-\frac{(m_gr)^2}{8}}- 40.0(4.3)\Lambda e^{-\frac{(m_gr)^2}{2}}.
\end{align}

Our current analysis shows that the Gaussian potential better matches the original data, as we see in Fig. \ref{fig:NBS_final}. We use a combination of bands from several lattice sets for each potential, helping us to find the final coefficients in these potentials. Through  the relation between the Yukawa potential's coefficient $Y$ and the coupling coefficient $\lambda_3$ in the scalar glueball Lagrangian (see Eq.(\ref{eq:lambda3_Y})), we find $\lambda_3=2m_g\sqrt{Y}=53.3(5.8)\Lambda$.

Because of these findings, we decide to use both potentials in our cross section results. We treat the difference between these two types of potential as a way to estimate the systematic uncertainties.

Having established the values of $Y$,  $Z^{(1)}$, and $Z^{(2)}$, we are now equipped to formulate the Schrödinger equation for the radial wave function $\Psi(r)$, which is related to $r\Psi_g(r)$, in the presence of a non-zero relative momentum $k$. This equation is expressed as:
\begin{align}
\left(\frac{d^2}{dr^2} + k^2 - m_gV(r)\right)\Psi(r) = 0.
\end{align}
When $r\to\infty$, $V(r)\to0$. It suggests that $\Psi(r)$ approaches an asymptotic form $\Psi(r) \propto \sin(kr + \delta(k))$ at large distances. Utilizing the Runge-Kutta method, we can accurately solve for $\Psi(r)$. Applying the parameters derived from the Yukawa and Gaussian potentials, we calculate the $s$-wave total cross sections separately. The results with only statistic uncertainties are as follows:
\begin{gather}
\sigma_{\mathrm{Yukawa}} = 1.110(15)\Lambda^{-2},\\
\sigma_{\mathrm{Gaussian}} = 3.716(52)\Lambda^{-2}.
\end{gather}

Our final result for the cross section is given as: 
\begin{align}
\sigma = (1.09\sim3.77)\Lambda^{-2},
\end{align}
which contains both statistic and  systematic uncertainties.

\begin{figure}
\centering
\includegraphics[scale=0.7]{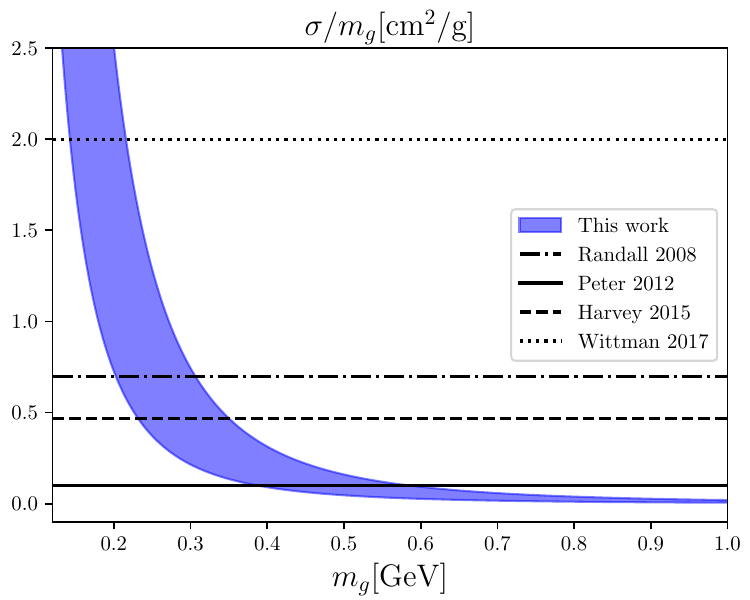}
\caption{This graph shows the ratio $\sigma/m_g$ against the dark glueball mass $m_g$. It incorporates the constraints  from several research groups that  have reported on the cross section of self-interacting dark matter. The band in the graph represents the range encompassed by our calculations. For context and comparison, the graph also includes results from established literature: $\sigma/m=0.7\mathrm{cm}^2/\mathrm{g}$ as in Ref.~\cite{Randall:2008ppe}, $\sigma/m=0.1\mathrm{cm}^2/\mathrm{g}$ from Ref.~\cite{Peter:2012jh}, $\sigma/m=0.47\mathrm{cm}^2/\mathrm{g}$ from Ref.~\cite{Harvey:2015hha}, and $\sigma/m=2.0\mathrm{cm}^2/\mathrm{g}$ from Ref.~\cite{Wittman:2017gxn}.}
\label{fig:sigma_m_m}
\end{figure}

Given our finding that both $\sigma$ and $m_g$ vary with $\Lambda$, we are able to use $m_g$ instead of $\Lambda$ when representing $\sigma/m$. This allows us to match our $\sigma/m$ result with experimental data and assign an appropriate value to $m$. Fig. \ref{fig:sigma_m_m} presents our findings alongside data from experimental studies \cite{Randall:2008ppe,Peter:2012jh,Harvey:2015hha,Wittman:2017gxn}. It is evident from our analysis that the ratio $\sigma/m_g$ tends to decrease as $m_g$ increases. Additionally, according to current constraint, it is estimated that the value of $m_g$ is likely around 0.3 GeV.

\section{Summary and outlook}

In this work, we have conducted an analysis of $0^{++}$ glueballs within the $SU(2)$ gauge theory using both Monte Carlo simulations and machine learning techniques.  We have used these two approaches to generate various  lattice configurations, and accordingly extracted the dark glueball mass using these configurations based on naive and improved actions.    
Using a coupling constant of $\beta=2.2$ as an illustration, we have compared the dark glueball mass calculated from the configurations generated from the two methods. While consistent results can be achieved, the two methods demonstrate distinct advantages. Given these considerations, our primary simulations have employed the Monte Carlo approach, while machine learning is only utilized for a comparative analysis.

Then the glueball interaction potential, crucial for extracting the interaction  coupling constant in effective quantum field theory and determining the glueball scattering cross section, is extracted. We then established  a connection  between the scattering cross section and the dark glueball mass, a vital aspect for understanding dark glueball behavior in various physical scenarios. Our findings show that the ratio $\sigma/m_g$ decreases with increasing $m_g$, aligning with values determined in experimental studies. This correlation, along with estimated dark glueball mass $m_g\approx0.3\;\mathrm{GeV}$ from experimental data, has significant implications for dark matter research.

The application  of Lattice simulations to dark glueball  not only showcases the potential of computational advancements in theoretical physics but also paves the way for more refined and efficient research methodologies in the future. Our results can contribute to the broader understanding of dark glueball dynamics and their interactions in the universe~\cite{Kang:2022jbg}, particularly in relation to dark matter. This study  encourages  further exploration into complex particle systems and their interactions, which could be pivotal in understanding the mysteries of dark matter and its fundamental forces. 

\section*{Acknowledgments}
We are very grateful to Ying Chen, Longcheng Gui, Zhaofeng Kang, Hang  Liu,  Jianglai Liu, Peng Sun,  Jinxin Tan,  Lingxiao Wang,  Yibo Yang, Haibo Yu, Ning Zhou, Jiang Zhu for valuable discussions. This work is supported in part by Natural Science Foundation of China under grant No. 12125503 and 12335003, and the Natural Science Foundation of Shandong province under the Grant No. ZR2022ZD26. The computations in this paper were run on the Siyuan-1 cluster supported by the Center for High Performance Computing at Shanghai Jiao Tong University and Advanced Computing East China Subcenter.

\appendix

\section{Lattice Spacing from Wilson Loop}
\label{sec:wiloop}
In the realm of lattice gauge theory, the expectation value of the Wilson loop is crucial. We use this value to set the lattice spacing for each lattice configuration. The static potential for the $SU(2)$ gauge field is determined as described in \cite{Bali:2000gf}:
\begin{align}
V(n_r) = \lim_{t\to\infty}\frac{1}{a}\ln\frac{W(n_r,n_t)}{W(n_r,n_t+1)},
\label{eq:V_WLP}
\end{align}
where $W(n_r,n_t)$ represents the Wilson loop's expectation value with length $n_r$ and width $n_t$ in lattice units. At short distances, this potential behaves as $1/n_r$, and at larger distances, it exhibits a linear growth, represented by the string tension $\sigma_{st}$. 

\begin{figure}
\centering
\includegraphics[scale=0.7]{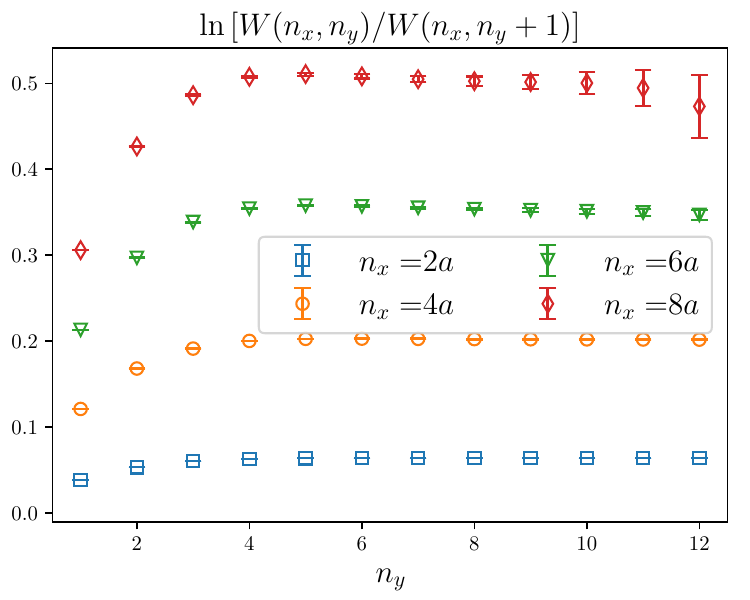}
\caption{The plot of $\ln\frac{W(n_x,n_y)}{W(n_x,n_y+1)}$ based on naive action with $\beta=2.4$.}
\label{fig:aV_plat}
\end{figure}

In our study, we focus on the space-space Wilson loop in $x-y$ plane. To represent $aV(n_r)$, we use the average value of $\ln[W(n_x,n_y)/W(n_x,n_y+1)]$ over a large range of $t$ values. As an illustration, we consider a basic setup with $\beta=2.4$, depicted in Fig. \ref{fig:aV_plat}. Here, we calculate $aV(n_x)$ as follows:
\begin{align}
aV(n_x)=\frac{1}{2}\left[\ln\frac{W(n_x,8)}{W(n_x,9)}+\ln\frac{W(n_x,9)}{W(n_x,10)}\right].
\end{align}
As shown in the figure, when $n_y \geq 6$, the term $\ln[W(n_x,n_y)/W(n_x,n_y+1)]$ stabilizes, forming a plateau. This supports our approach is conservative.

\begin{figure}
\centering
\includegraphics[scale=0.7]{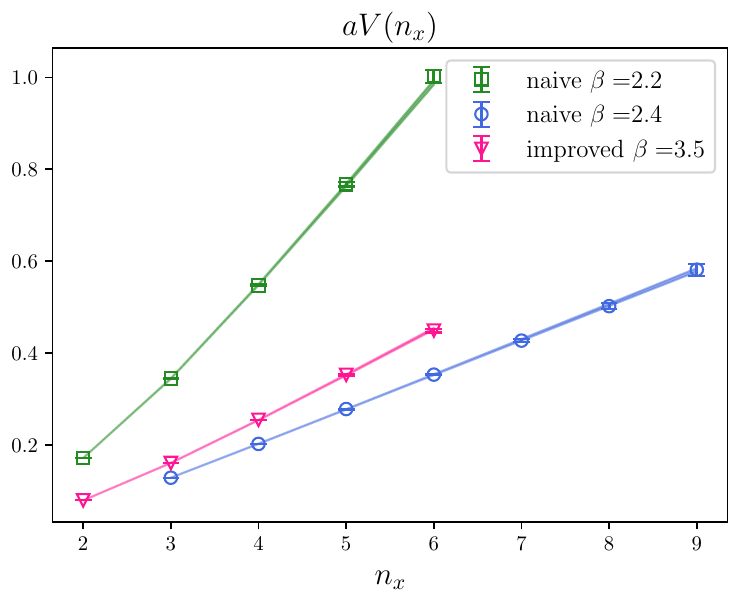}
\caption{Fitting results for $aV(n_x)$. The string tension $a^2\sigma_{st}$ for cases naive $\beta=2.2$, $2.4$ and improved $\beta=3.5$ are $0.2359(29)$, $0.0766(13)$, $0.1053(11)$.}
\label{fig:aV_fit}
\end{figure}

\begin{figure}
\centering
\includegraphics[scale=0.7]{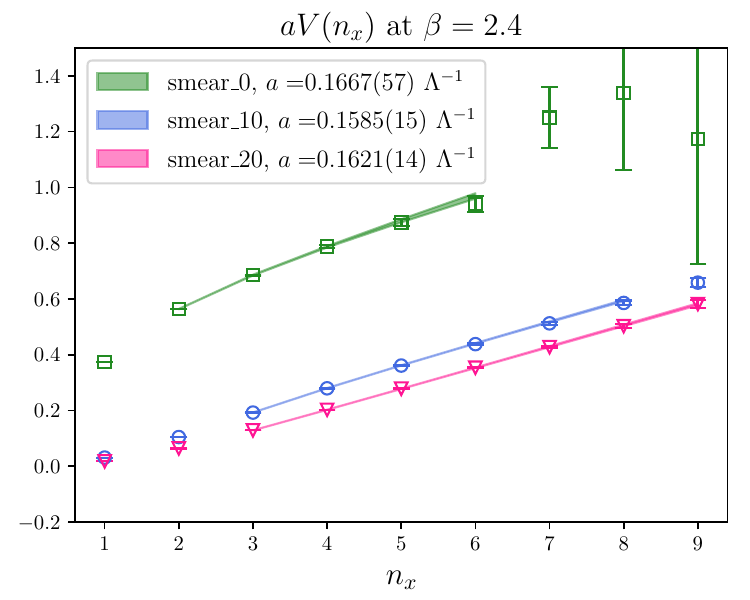}
\caption{Static potentials from configurations with or without APE smearing at $\beta=2.4$. The number in the label after "smear\_" represents times of smearing.}
\label{fig:aV_sm}
\end{figure}

One can express $aV(n_x)$ as a function of $n_x$:
\begin{align}
aV(n_x) = A+\frac{B}{n_x}+a^2\sigma_{st} n_x.
\label{eq:aV}
\end{align}
For each lattice setup, we fit the Wilson loop data to this formula to obtain the string tension $a^2\sigma_{st}$. The outcomes of our fitting process are illustrated in Fig. \ref{fig:aV_fit}, where we have selected examples with naive $\beta=2.2$ and $2.4$, alongside an improved $\beta=3.5$ for demonstration. Fig. \ref{fig:aV_fit} reveals that the formula in Eq.(\ref{eq:aV}) effectively captures the string tension derived from our Wilson loop data.  It should be noted that in the small spatial  separation  region, the $B$ term will be influenced by APE smearing, resulting in the modification of the sign of $B$. However, the string tension is not significantly affected by smearing. The comparison of static potential without and with 10/20 times  APE smearing is shown in Fig. \ref{fig:aV_sm}, supporting the above observation.  The effect on the lattice spacing is about $(1\sim 2 )\sigma$. 

Using the relationship between string tension and the energy scale $\Lambda$ for $SU(2)$ gauge theory~\cite{Teper:1998kw}:
\begin{align}
\Lambda = 0.586(45)\sqrt{\sigma_{st}},
\end{align}
we finally determine the lattice spacing in units of $\Lambda$. The results are displayed in Tab. \ref{tab:lattice_spacing}.

\section{Determining Wave Function from a Specified Potential}
In Sec. \ref{sec:scatter}, we outline our method to determine coefficients for a specific potential using the radial wave function $R(r)$, satisfying:
\begin{align}
\frac{d^2R(r)}{dr^2}+\frac{2}{r}\frac{dR(r)}{dr}-m_gV(r)R(r)=0.
\end{align}
Here, $V(r)$ may represent Yukawa or Gaussian potentials, each requiring one or two coefficients. Our approach is to use a test coefficient in $V(r)$ and solve the differential equation via the Runge-Kutta method, starting from $R(r_{\mathrm{ini}})$ and $R'(r_{\mathrm{ini}})$. The initial condition is derived from the NBS wave function of lattice data, replacing differentiation with difference. 

Applying the fourth-order Runge-Kutta method, we rewrite the above equation as:
\begin{align}
R''(r)=f(r,R,R')=\frac{2}{r}R'(r)-m_gR(r)V(r).
\end{align}
Then we can obtain the solution $R(r)$ by iterating the following equations with step length $h$:
\begin{align}
&k_1R = hr,\;\;k_1R' = hf(r,R,R'),\\
&k_2R = h(R'+\frac{1}{2}k_1R'),\\
&k_2R' = hf(r+\frac{1}{2}h,R+\frac{1}{2}k_1R,R'+\frac{1}{2}k_1R'),\\
&k_3R = h(R'+\frac{1}{2}k_2R'),\\
&k_3R' = hf(r+\frac{1}{2}h,R+\frac{1}{2}k_2R,R'+\frac{1}{2}k_2R'),\\
&k_4R = h(R'+k_3R'),\\
&k_4R' = hf(r+h,R+k_3R,R'+k_3R'),\\
&r_1 = r+h,\\
&R_1 = R+\frac{1}{6}(k_1R+2k_2R+2k_3R+k_4R),\\
&R'_1 = R'+\frac{1}{6}(k_1R'+2k_2R'+2k_3R'+k_4R'),
\end{align}
where $r_1$, $R_1$, and $R'_1$ are the iteratively updated values at $r+h$. Using test coefficients in $V(r)$, we approximate the corresponding solution, denoted as $R_{\mathrm{test}}(r)$. We then assess the distance between this test solution and our lattice data. Then by varying the coefficients within a certain range and finding the minimal distance, we eventually identify the corresponding coefficients for $V(r)$.

\end{document}